# Similar Modification of Intermittent Density Bursts by Electrode Biasing and Dynamic Ergodic Divertor in the TEXTOR Tokamak

Irakli S. Nanobashvili[1,2]

[1]*Andronikashvili Institute of Physics, Iv. Javakhishvili Tbilisi State University, Tamarashvili str. 6, 0177 Tbilisi, Georgia*
[2]*Institute of Theoretical Physics, Ilia State University, Kakutsa Cholokashvili ave. 3/5, 0162 Tbilisi, Georgia*
e-mail address: inanob@yahoo.com

Intermittent positive bursts of plasma density detected by Langmuir probes at the edge of the TEXTOR tokamak are investigated. Burst statistical properties and temporal characteristics together with their radial dependence are studied in two different regimes – with electrode biasing and dynamic ergodic divertor (DED). Similar modification of intermittent burst characteristics are observed. Namely, the average burst rate increases and the average burst duration decreases compared to Ohmic conditions. Statistical properties are also modified in the same way in both regimes. The reason is that biasing and certain regimes of DED cause the modification of radial electric field which affects similarly the dynamics of coherent turbulent structures and plasma transport through $E_r \times B_t$ induced sheared poloidal rotation. Thus, after detailed investigations and refinement, certain regimes of DED can be used as "contactless biasing" for the external control of plasma turbulent transport in fusion devices.

I. INTRODUCTION

Investigation of plasma turbulent transport at the edge of tokamaks is one of the most important and interesting tasks of modern fusion research. The transport is highly bursty, intermittent and has a strongly convective character. Large turbulent events – density bursts bring important contribution to such transport. Density bursts are formed intermittently on diffusive background and propagate radially outwards at a speed which is a fraction of ion sound speed. This can result in degradation of confinement, strong erosion and heat load on first wall and other plasma facing components together with unwanted retention of tritium. It is very important to understand the physical nature of bursty turbulent transport in general and especially in the context of developing the methods and tools for its external control.

Investigation of temporal characteristics of intermittent density bursts such as burst rate, inter-burst time and burst duration together with their statistical properties is an efficient method for better understanding of turbulent transport in tokamak edge plasma [1-4]. In the present paper we report the results of such investigation of intermittent density bursts measured at the edge of the TEXTOR tokamak [5,6] by means of reciprocating Langmuir probe in two different regimes – with electrode biasing and dynamic ergodic divertor (DED).

On the TEXTOR tokamak two methods are used for external control of plasma turbulent transport, namely electrode biasing [5] and the DED [6-8]. Both have strong influence on edge plasma transport and electrode biasing can even trigger the transition from low to high confinement mode [5].

II. EXPERIMENTAL SETUP

During the analyzed biasing discharges of TEXTOR (major radius *R = 1.75 m* and minor radius *a = 0.475 m*), the plasma current was $I_p$ = *200 kA*, the toroidal magnetic field $B_t = 1.9\ T$, the line average density of plasma $1 \times 10^{19}\ m^{-3}$, and the biasing voltage $V_{bias} = 150\ V$. In order to bias the TEXTOR plasma canoe-shaped electrode is inserted in it. The biasing voltage is applied between the electrode and toroidal belt limiter (which is grounded to the liner) in the stationary Ohmic phase of the discharge. A first radial scan with the probe is made during the Ohmic phase, and a second one during the biasing phase. The probe head is installed at the equatorial plane on low-field side (LFS) of the torus and consists of seven pins. Ion saturation current $I_{sat}$, floating potential $V_{fl}$ and electron temperature $T_e$ were measured (part of the pins were set-up as a triple probe) in the same way as described in the papers [7,8]. The sampling frequency is 500 kHz. The radial electric field is obtained from the derivative of the plasma potential $V_{pl} = V_{fl} + 2.5 T_e$ [9]. The same measurements have been performed also during discharge with DED for which the conditions were the following: $I_p = 250\ kA$, $B_t = 2.25\ T$, the line average density of plasma $1.5 \times 10^{19}\ m^{-3}$, DED current $I_{DED} = 3\ kA$. On the TEXTOR tokamak the DED principally consists of sixteen magnetic perturbation coils mounted in the vacuum vessel on high field side of the torus. The coils are helically wound and parallel to the field lines on the magnetic flux surface the safety factor of which equals to 3. The current can be distributed in different ways in the DED coils and the base poloidal/toroidal modes 12/4, 6/2 and 3/1 can be obtained. The DED current is applied in the stationary Ohmic phase of the discharge. For the discharge under study the base poloidal/toroidal mode 6/2 is used. A first radial scan with the probe is made during Ohmic phase, and a second one during DED.

III. MAIN CONSIDERATION

In the Ohmic phase of the biasing discharge we observe intermittent positive bursts of the ion saturation current $I_{sat}$ in all radial positions, but as the radial distance increases they are more dominant in the signal. This is confirmed by statistical analysis – the probability density function (PDF) of $I_{sat}$ is positively skewed and skewness increases radially (see Fig.1).

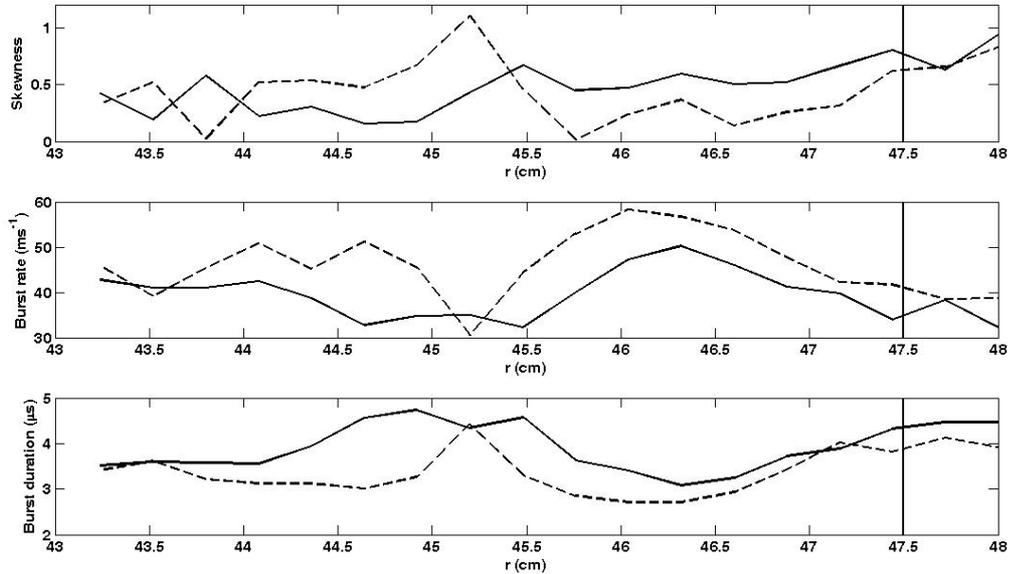

Fig.1 *Radial dependence of $I_{sat}$ (a) skewness, (b) average burst rate, and (c) average burst duration before (solid lines) and during (dashed lines) biasing phase of the TEXTOR discharge #112172. Vertical line shows the radial position of the limiter.*

A similar radial dependence of $I_{sat}$ skewness has been reported in the papers [1-4,10-12]. Fig.1 shows that during biasing the skewness rises in 1.5 cm wide radial region starting at 44 cm. At larger radii $I_{sat}$ skewness decreases compared to the Ohmic phase. The radial dependence of the temporal characteristics of intermittent bursts during the biasing discharge is also presented in the Fig.1. During biasing the average burst rate increases and the average burst duration decreases compared to the Ohmic phase. It must be mentioned that a similar modification of intermittent burst temporal characteristics has been already observed on the CASTOR tokamak [1]. The reason of such modification is that biasing generates the strongly nonuniform radial electric field, changes the radial electric field which already existed in Ohmic phase (see the Fig.2) and imposes stronger sheared poloidal rotation on the plasma.

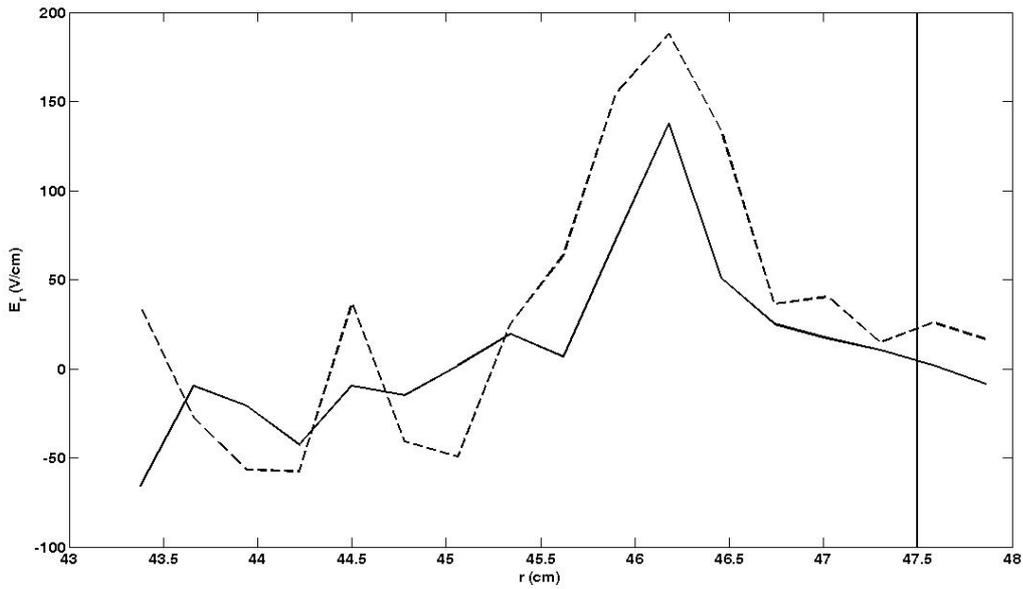

Fig.2 *Profile of the radial electric field before (solid lines) and during (dashed lines) biasing phase of the TEXTOR discharge #112172. Vertical line shows the radial position of the limiter.*

The sheared poloidal rotation splits coherent structures, which are responsible for the appearance of intermittent bursts [12], into smaller structures and moves them faster in poloidal direction [1]. As a result the Langmuir probe detects more bursts in biasing phase of the discharge and their average duration decreases [1].

Generally, during DED operation open stochastic magnetic field lines appear in plasma boundary and radial magnetic connection between the edge plasma and wall is created. Electrons move faster than ions along the radial field lines to the wall. As a result the radial electric field is modified in the plasma [7,8]. Since the radial electric field is modified during DED operation and at the same time different DED regimes have different influence on plasma [7], one can presume that in a certain DED regime we may get the similar modification of intermittent bursts as in the case of electrode biasing. Indeed, we found such DED regime among many different ones used on TEXTOR. During DED phase of such discharge the $I_{sat}$ skewness rises in a 1.5 cm wide radial region and at larger radii it drops compared to the Ohmic phase (see the Fig.3). As for the intermittent burst temporal characteristics, average burst rate increases and average burst duration decreases compared to the Ohmic phase (see the Fig.3). These modifications are quite similar to those observed during electrode biasing.

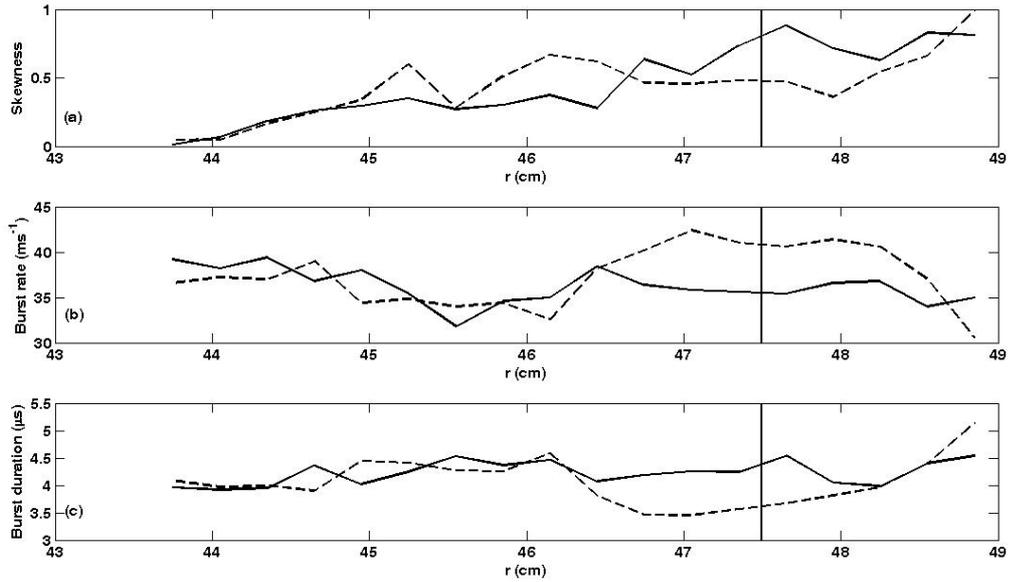

Fig.3 *Radial dependence of $I_{sat}$ (a) skewness, (b) average burst rate and (c) average burst duration before (solid lines) and during (dashed lines) DED phase of the TEXTOR discharge #111626. Vertical line shows the radial position of the limiter.*

The main reason for the above mentioned changes should be the modification of the radial electric field by DED which is presented in the Fig.4.

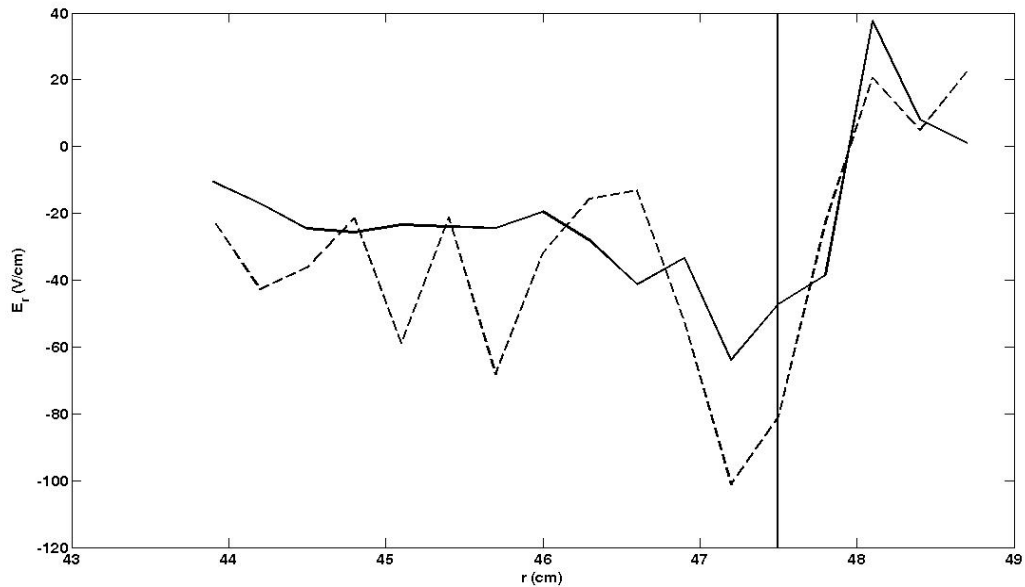

Fig.4 *Profile of the radial electric field before (solid lines) and during (dashed lines) DED phase of the TEXTOR discharge #111626. Vertical line shows the radial position of the limiter.*

On the Fig.4 one can see that the radial position of the shear layer practically does not change in the DED regime, but the radial electric field is stronger (more negative) inside this layer compared to Ohmic conditions. The electric field slightly decreases outside the shear layer, but the difference between the peaks of the field (which also do not change their radial location) around the shear layer increases during DED. Thus, the sheared poloidal rotation is stronger during DED. As a result coherent structures are splitted into smaller structures which move faster in poloidal direction and intermittent burst characteristics are modified in the same way as during electrode biasing regime.

IV. CONCLUSION

In conclusion, we report the study of intermittent burst characteristics at the edge of the TEXTOR tokamak and their modification during electrode biasing and DED operation. Similar modifications are observed in two regimes. The reason is that these regimes modify the radial electric field which affects similarly the dynamics of coherent turbulent structures and plasma transport through $E_r \times B_t$ induced sheared poloidal rotation. Thus, after detailed investigations and refinement, certain regimes of DED can be used as "contactless biasing" which should be beneficial for the external control of plasma turbulent transport in next generation fusion devices.

V. ACKNOWLEDGEMENTS

This work was carried out during authors visit to Forschungszentrum Jülich (Germany), which was supported by the Erasmus Mundus Higher Education Program.

The work was partially supported by Shota Rustaveli National Science Foundation (Georgia) Grant FR/443/6-140/11.

Useful discussions of the problem with Professor Guido Van Oost are gratefully acknowledged.

The author thanks Y. Xu and I. Shesterikov for supplying the TEXTOR probe data.